# Simulating the Transverse Ising Model on a Quantum Computer: Error Correction with the Surface Code

Hao You (尤浩)[1,2], Michael R. Geller[1], P. C. Stancil[1,2]
1. *Department of Physics and Astronomy, University of Georgia, Athens, GA 30602*
2. *Center for Simulational Physics, University of Georgia, Athens, GA 30602*

We estimate the resource requirements for the quantum simulation of the ground state energy of the one-dimensional quantum transverse Ising model based on the surface code implementation of a fault-tolerant quantum computer. The surface code approach has one of the highest known tolerable error rates (~1%) which makes it currently one of the most practical quantum computing schemes. Compared to results of the same model using the concatenated Steane code, the current results indicate that the simulation time is comparable but the number of physical qubits for the surface code is 1-2 orders of magnitude larger than that of the concatenation code. Considering that the error threshold requirements of the surface code is four orders of magnitude higher than the concatenation code, building a quantum computer with a surface code implementation appears more promising given current physical hardware capabilities.

## I. INTRODUCTION

Since Feynman proposed that it may be possible to use a quantum system for simulation of the basic properties of another quantum system [1], many different approaches for implementing quantum simulation have been investigated [2-11]. In these early studies, resource requirements of the quantum computer were estimated without considering the effects of decoherence. Later, quantum error correction and fault-tolerant quantum computation were incorporated into resource estimation, for example, for the quantum simulation of the ground state energy in the transverse Ising model (TIM) [12], there is a large number of physical qubits and lengthy computational time required for the quantum computation scheme with error correction via the concatenation code [13-19].

In addition to the concatenation code implementation, a class of approaches to building a quantum computer with the topological code (which is a kind of stabilizer code [20]) has been proposed [21-37]. Topological error-correction codes store the quantum information by associating it with some topological properties of the system and thus have the highest known tolerable error rate. Among all the topological codes, the surface code [21-22,26-28,36] is currently considered to be one of the most practical fault-tolerant quantum computing schemes because the operation time and the resource overhead are within reasonable limits [38].

Here we investigate the quantum simulation of the TIM ground state energy on a surface code quantum computer. We have chosen to investigate this model since it is well studied with the concatenation code [12]. Our main interest is to understand connections of the resource requirements between the surface code and concatenation code. Both of these approaches are stabilizer codes [20], however they are quite different. For this purpose, we use the same quantum algorithm as that in Ref. [12] and estimate the number of physical qubits and the computational time for the simulation of the TIM.

Section II gives an overview of the operation of the surface code. Section III maps the calculation of the ground state energy for the TIM onto a quantum phase estimation circuit that includes the effects of surface code quantum error correction. The estimation of the number of physical qubits and the computational time are then presented. A summary and conclusions are given in Section IV.

## II. QUANTUM COMPUTING WITH THE SURFACE CODE

The surface code is a stabilizer code associated with a two-dimensional square lattice of physical qubits [21-22,26-28,36]. One of the significant advantages of the surface code is that error tolerances using only one- and two-qubit nearest-neighbor gates results in an error threshold of somewhat less than 1% per gate [22,26-28]. As this results in a significant relaxation on the physical performance requirements for a single qubit, a number of schemes for physical implementation of surface code quantum computation have been proposed, using superconductors [36,39,40] and semiconductor nanophotonics [41,42]. An equivalent version of the one-way quantum computer using a three-dimensional cluster state [26,27] could be implemented using ion trap [43] and photonic approaches [44,45]. Recently the first experimental demonstration of topological error correction with a photonic cluster state was reported [46,47].

A representation of scalable logical qubits for the surface code is illustrated schematically in Fig. 1. There are two kinds of physical qubits in the square lattice: data qubits indicated by filled circles and measurement qubits indicated by open circles. In order to implement surface code error correction, all the data and measurement qubits must perform the following physical operations: state initialization, measurement of the qubit along the Z axis, single-qubit rotation, a two-qubit controlled-NOT between the nearest neighbors and a swap operation between the data and measurement (syndrome) qubits. The computational basis states are coded in the data qubits. The measurement qubits are used to read the quantum state of the data qubits and project into eigenstates of stabilizers. For topological clarification, stabilizers are represented by colored four-terminal stars in Fig. 1: a tensor product of Z operators (Z-stabilizer) colored green and a tensor product of X operators (X-stabilizer) colored in yellow. Individual measurement qubits are operated in such a way that each project four neighboring data qubits onto an eigenstate of the Z-stabilizer $\hat{Z}_{1234}$ or X-stabilizer $\hat{X}_{1234}$. Thus operation of the measurement qubits project all the data qubits on the surface into a quiescent state [36] which is a simultaneous eigenstate of all the stabilizers $\hat{Z}_{1234}$ and $\hat{X}_{1234}$. One round of such measurements plus possibly other computational steps on the surface is defined as a surface code cycle. As can be seen in Figures 1b and c, one surface code cycle is comprised of 8 physical steps [36].

Errors occurring on the surface code induce sign flips on the reported eigenvalues of the given stabilizers. To cope with these errors, one repeats the surface code cycle, keeping track of the reported eigenvalues of each stabilizer change. If errors are sufficiently rare, the error syndromes can be efficiently matched to deduce on which qubit the error occurred with this being handled by control software on a classical computer [28,36,48-51].

Each logical qubit is comprised of two smooth defects where Z-stabilizers in the defects are turned off. This kind of logical qubit is called a smooth qubit. (In fact, there is another kind of logical qubit, the rough qubit (not shown), which is comprised of two rough defects where X-stabilizers in the defects are turned off.) As can be seen in Fig. 1, for a smooth qubit $Q$, the logical Pauli-Z operator $\hat{Z}_L$ is the ring of local Z operators around one defect. The logical Pauli-X operator $\hat{X}_L$ is the chain of local X operators connecting the two defects. The code distance $d$ in the surface code is defined as the length of the shortest logical operator chain. In the configuration of Fig. 1, a product of Pauli operators of seven data qubits is the minimum needed to change the state of the logical qubit so that the code distance $d$ is seven. Scalable logical qubits are equally spaced by $d$. In addition to spatial separation $d$, logical operations should also be separated in the time domain by around $d$ surface code cycles. Thus code distance determines both the size of surface code quantum computer and the time duration of logical operations [36].

For a given hardware architecture, the single-step physical error probability $p$ is assumed to be known. The logical error rate per surface code cycle $p_L$ decreases with the value of $d$ as [36]:

$$p_L \approx 0.043(p/p_{th})^{(d+1)/2}, \tag{1}$$

where the error threshold error rate $p_{th} = 0.57\%$.

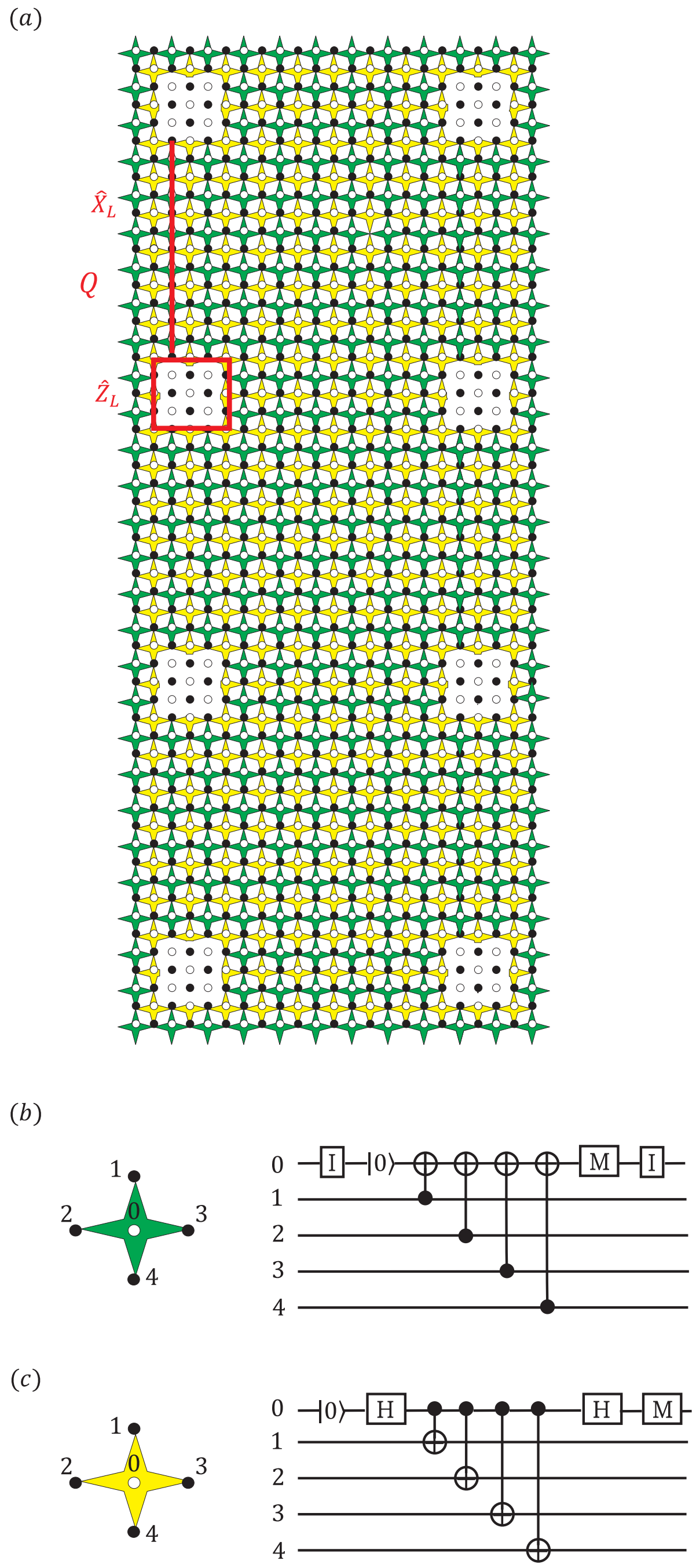

Fig. 1. (a) Two-dimensional array implementation of the surface code. Filled circles represent data qubits and open circles represent measurement qubits. Stars in green represent Z-stabilizers and stars in yellow represent X-stabilizers. The logical qubit is composed of a double-defect in Z-stablizers. The logical Pauli operators for the logical qubit $Q$ are indicated in red. $\hat{Z}_L$ is the product of Pauli-Z operators of 8 data qubits on the ring in red and $\hat{X}_L$ is the product of Pauli-X operators of 7 data qubits on the chain in red. The code distance of this configuration is 7 and all logical qubits are separated with this code distance. Four logical qubits are shown in the figure. The sides of this two-dimensional array extend outwards where more logical qubits are located. (b) Representation of Z-stabilizer $\hat{Z}_{1234} = \hat{Z}_1\hat{Z}_2\hat{Z}_3\hat{Z}_4$ in the surface code (left) and quantum circuit (right) for one surface code cycle for measurement of $\hat{Z}_{1234}$. The Z-stabilizer is indicated by a star in green. Data qubits 1, 2, 3 and 4 are located at terminals of the star and the measure qubit 0 is located at the center of the star. After initialization of qubit 0, stabilizer measurement includes four CNOT operations followed by a projective measurement of qubit 0. (c) Representation of X-stabilizer $\hat{X}_{1234} = \hat{X}_1\hat{X}_2\hat{X}_3\hat{X}_4$ in the surface code (left) and quantum circuit (right) for one surface code cycle for measurement of $\hat{X}_{1234}$. The X-stabilizer is indicated by a star in yellow. Data qubits 1, 2, 3 and 4 are located at terminals of the star and the measure qubit 0 is located in the center of the star. After initialization of qubit 0, the stabilizer measurement includes two Hadamard operations, four CNOT operations, and a projective measurement of qubit 0.

Instead of physical manipulations on the related data qubits, Pauli logical operations on a surface code quantum computer can be performed virtually in classical software [28,36]. The linearity of the quantum computational process allows these operators to be tracked in software, with their effects dealt with at the end of the computation. The Gottesman-Knill theorem states that the tracking can be done efficiently on a classical computer [52].

A CNOT gate on a surface code quantum computer is a topological fault-tolerant gate [26-28,36]. Using a smooth qubit as control and a rough qubit as target, a CNOT gate can be generated by braiding the smooth qubit around the rough qubit. This implementation allows a single-control-bit multiple-target CNOT gate operation in the same amount of time as a single CNOT.

However, this implementation of a CNOT is limited as the control must always be a smooth qubit and the target must always be a rough qubit. In order to obtain a CNOT between arbitrary types of qubits, the surface topology has to be changed [26,27]. The implementation for a CNOT between two smooth qubits is illustrated in Fig. 2. In addition to the control and target qubits, two ancillas $a_1$ and $a_2$ are used to assist the operation. All qubits in the circuit are smooth qubits except the rough qubit $a_1$. Three logical CNOTs between smooth and rough qubits are generated by braiding between "target in" and $a_1$, "control in" and $a_1$, and finally $a_1$ and $a_2$. After measurements on target in and $a_1$, the net effect in the circuit is that a CNOT operation is generated between two smooth qubits.

Because stabilizations of the stabilizer values need to be protected by sufficient temporal distance [36], it could take $2.5d$ surface code cycles for the implementation of a generic CNOT gate between two smooth qubits [38,53]. This is the minimum volume logical CNOT. The volume per unit time is 6 (in units of $1.25d$), corresponding to a patch with 3 double-defect smooth qubits [38].

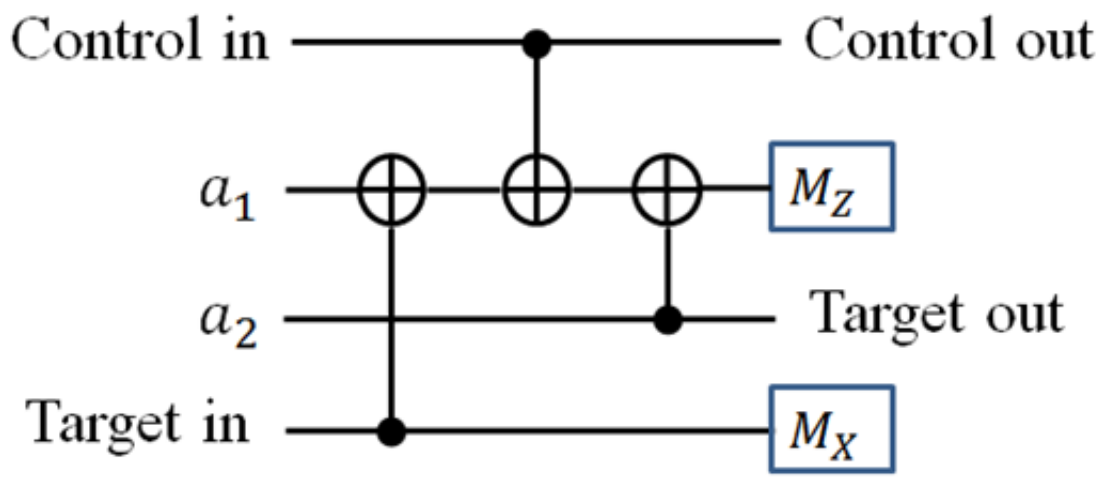

Fig. 2. Implementation to perform a CNOT gate between two smooth qubits [36]. The Control in, Target in and ancilla $a_2$ are all smooth qubits. Ancilla $a_1$ is a rough qubit. Braiding operations are used to generate logical CNOTs between smooth and rough qubits. After three braiding operations, measurements $M_Z$ on $a_1$ and $M_X$ on Target in are performed. The measurement outcomes are used to interpret the output states.

A Hadamard gate in the surface code can be generated by the Hadamard-double swap process [36,54]. In order to detect and decode errors in the time domain, waiting operations, similar to that for a CNOT gate, are necessary. It is convenient to expand the duration of a Hadamard gate in $2.5d$ surface code cycles [38,53].

To complete the universal set of quantum computation operations, surface code implementations of S and T gates are sufficient [26-28,36]. The single-qubit S and T gates are represented by matrices:

$$\hat{S} = \begin{pmatrix} 1 & 0 \\ 0 & i \end{pmatrix} \qquad (2)$$

and

$$\hat{T} = \begin{pmatrix} 1 & 0 \\ 0 & e^{i\pi/4} \end{pmatrix}. \qquad (3)$$

Logical implementation of each of these gates in the surface code is based on special ancilla states: implementing the S gate requires the $|Y_L\rangle$ ancilla state

$$|Y_L\rangle = \frac{1}{\sqrt{2}}(|0_L\rangle + i|1_L\rangle). \qquad (4)$$

While implementing the T gate requires the $|A_L\rangle$ ancilla state

$$|A_L\rangle = \frac{1}{\sqrt{2}}(|0_L\rangle + e^{i\pi/4}|1_L\rangle). \qquad (5)$$

These logical states are initialized non-fault-tolerantly and then distilled into high fidelity states by state distillation [55,56]. After the distillation, these high fidelity states are fed into logical circuits involving CNOT and Hadamard gates, shown in Fig. 3 and Fig. 4, respectively.

The circuit for the S gate implementation includes two logical CNOTs and two logical Hadamard operations [57,42,36]. An input state $|\psi_L\rangle$ is deterministically transformed into $S|\psi_L\rangle$. After the implementation, the $|Y_L\rangle$ ancilla state doesn't change and is re-useable for implementation of another S gate.

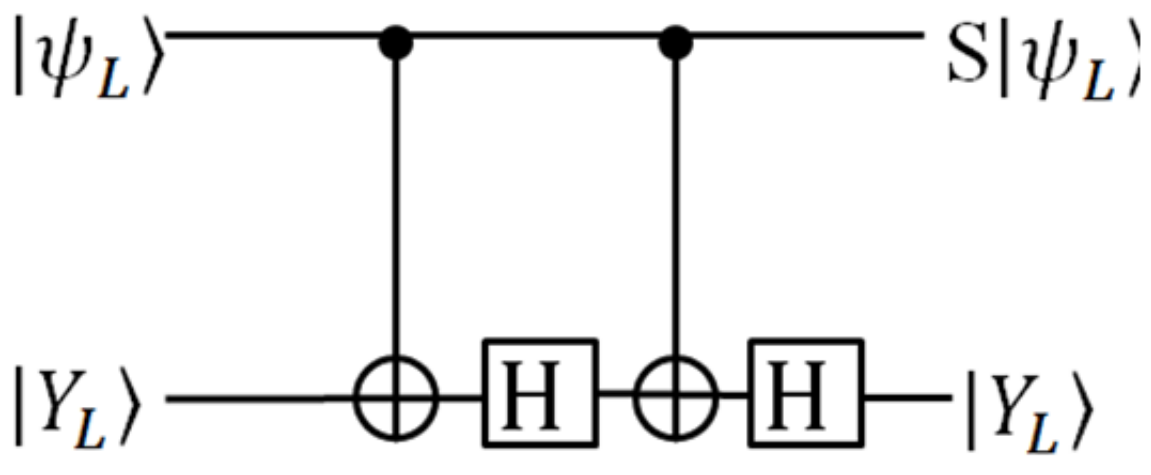

Fig. 3. The logic circuit that implements the S gate. The input state $|\psi_L\rangle$ is transformed to output state $S|\psi_L\rangle$.

The circuit for a T gate implementation is performed non-deterministically as shown in Fig. 4. Given an input state $|\psi_L\rangle$, the resulting state after a logical CNOT and a logical $M_z$ measurement depends on the readout of the $M_z$ measurement. If the measurement outcome is $M_z = 1$, the output state is $T|\psi_L\rangle$. However, if the measurement outcome is $M_z = -1$, then we must apply a $SX$ operation (Pauli X followed by an S gate) so that the output state is $T|\psi_L\rangle$. After the implementation, the $|A_L\rangle$ ancilla state is destroyed and is not re-useable. In order to implement another T gate, one must again initialize a $|A_L\rangle$ state non-fault-tolerantly and then distill into high fidelity states by state distillation.

The T and S gates can be completed in $11.25d$ and $10d$, respectively [53].

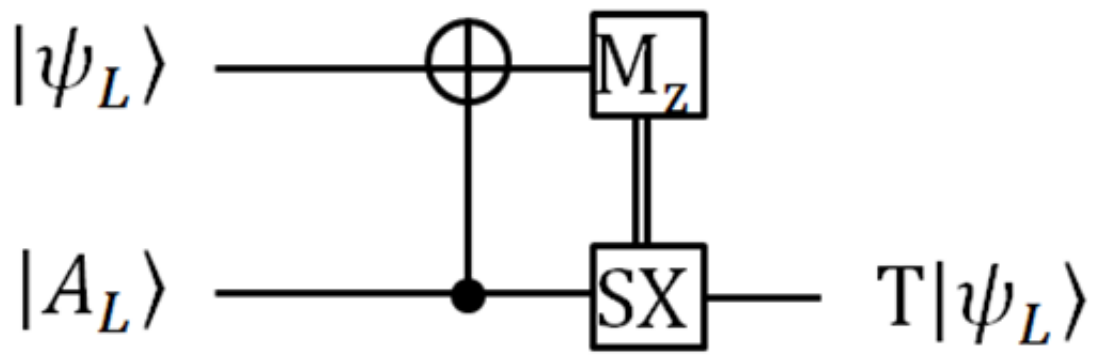

Fig. 4. The logic circuit that implements the T gate. The input state $|\psi_L\rangle$ is transformed to output state $T|\psi_L\rangle$.

As the $|Y_L\rangle$ ancilla state is re-useable, the distillation of the $|Y_L\rangle$ state only contributes to the resource requirements at the beginning of the simulation. The distillation of the $|A_L\rangle$ state is more critical as it cannot be re-used. Thus the distillation process for the $|A_L\rangle$ state is implemented concurrently with the simulation to anticipate the demands of the T gates. If the distillation of the $|A_L\rangle$ isn't completed when the T gates are required, it will affect the computational time and the number of physical qubits required for the simulation.

The distillation process of the $|A_L\rangle$ state is shown in Fig. 5 [36]. In this circuit, sixteen logical qubits serve as inputs, with the circuit generating one output state $|\psi_L\rangle$ and fifteen measurement outcomes. The circuit for the $T_L^\dagger$ gate can be found in Ref. [36] and it requires an ancilla qubit in the $|A_L\rangle$ state, which is prepared by the state injection or in a previous distillation round. The measurement patterns are used to determine whether the output state $|\psi_L\rangle$ is a good approximation to $|A_L\rangle$. The distillation converges rapidly to a nearly perfect output state $|A_L\rangle$. If the original states have error rates $p$, the output state will have an error rate $35p^3$ with exponential improvement. If one cycle of distillation does not result in a sufficiently accurate output, another cycle, level-2 distillation may be required. This results in an error rate $35^4p^9$.

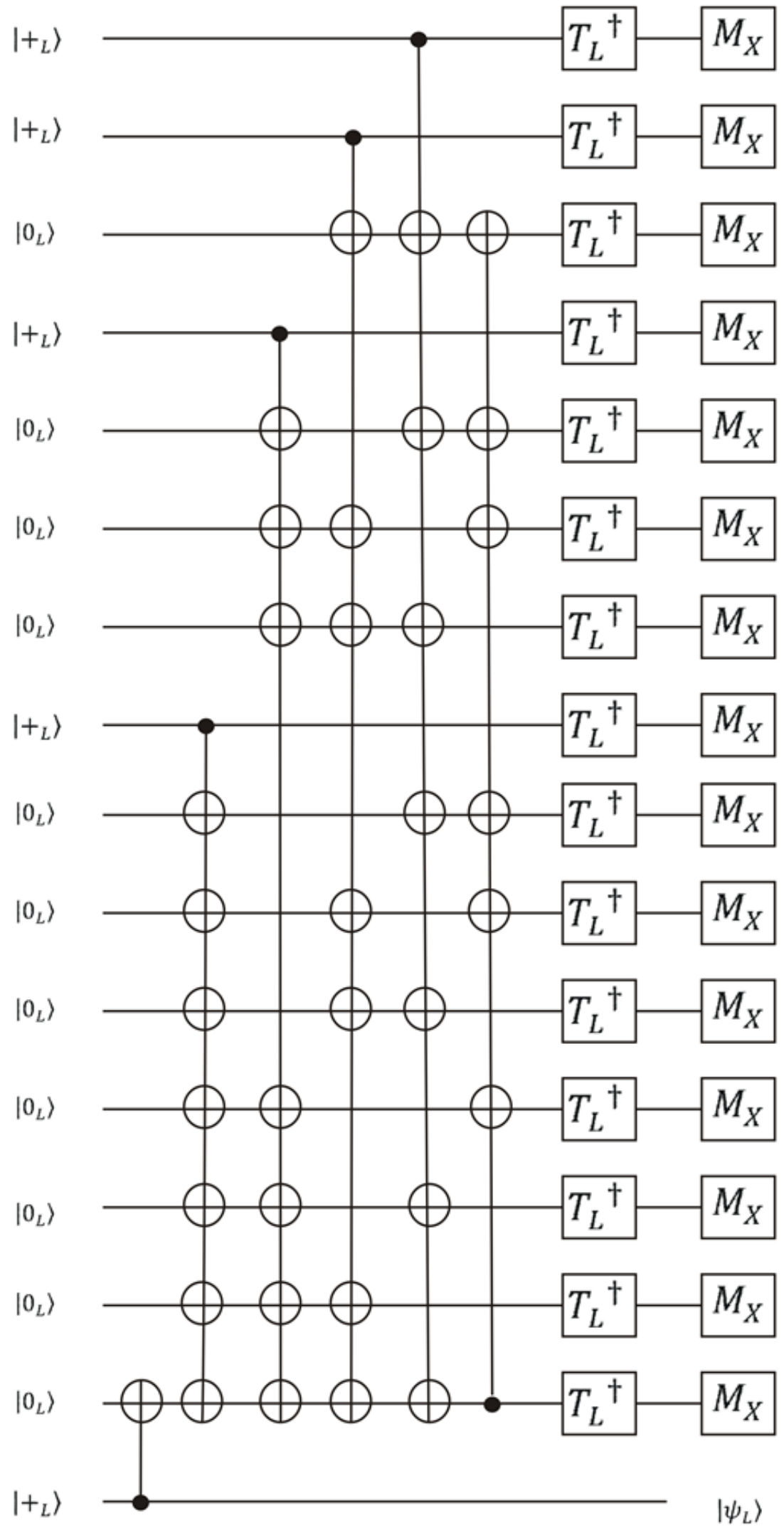

Fig. 5. Distillation circuit for the $|A_L\rangle$ state [36]. Sixteen logical qubits are run through this circuit, yielding the output state $|\psi_L\rangle$ to be $|A_L\rangle$ with higher fidelity.

For clarification, Table I summarizes the resource requirements for implementation of the basic set of gates with the surface code. These considerations are used in Section III to estimate the total resource requirements to compute the TIM ground state on a surface code quantum computer.

| Gate | Time (surface code cycles) | Ancilla (qubits) |
|---|---|---|
| CNOT | $2.5d$ | 1 |
| H | $2.5d$ | no |
| T | $11.25d$ | 1 un-recyclable, 1 reuse (for SX) |
| S | $10d$ | 1 re-usable |
| Pauli | instantaneous | no |

Table. 1. Basic set of logical gates for the surface code.

## III. TIM SIMULATION WITH THE SURFACE CODE

The TIM is an appropriate benchmark model for quantum simulation considering that it is one of the central problems in quantum phase transitions and quantum annealing. Further, quantum speedup over classical algorithms may be feasible for high spatial dimension TIM problems. In Ref. [12], Clark et al. focused on the quantum simulation of the ground state energy of the TIM for the one dimensional (1D) case with the concatenated Steane quantum error correcting code [58]. It is expected that the increase in resource requirements for a high-dimensional TIM problem will scale by a factor less than the problem spatial dimension [12].

In order to compare the effects of the surface code with that of the concatenation code for the quantum simulation of the TIM, we will utilize the same quantum algorithm adopted in Ref. [12] and estimate resource requirements for calculation of the ground state energy of the 1D TIM on a surface code quantum computer. The Hamiltonian of interest is:

$$H_I = -\sum_{j=1}^{N} X_j - \sum_{j=1}^{N-1} Z_j Z_{j+1}, \quad (6)$$

where $N$ is the number of spin-1/2 particles. $X_j$ and $Z_j$ are Pauli matrix operators.

Firstly, the problem of computing the eigenvalues of the Hamiltonian in Eq. (6) can be mapped onto the iterative phase estimation quantum circuit [59]. The phase estimation algorithm calculates an $M$-bit estimate of the phase $\phi$ of the eigenvalue $e^{-i2\pi\phi}$ of the time evolution unitary operator $U(\tau) = e^{-iH_I\tau}$ for a fixed time interval $\tau$, given an estimate of the ground state of $H_I$. The implementation is illustrated schematically in Fig. 6.

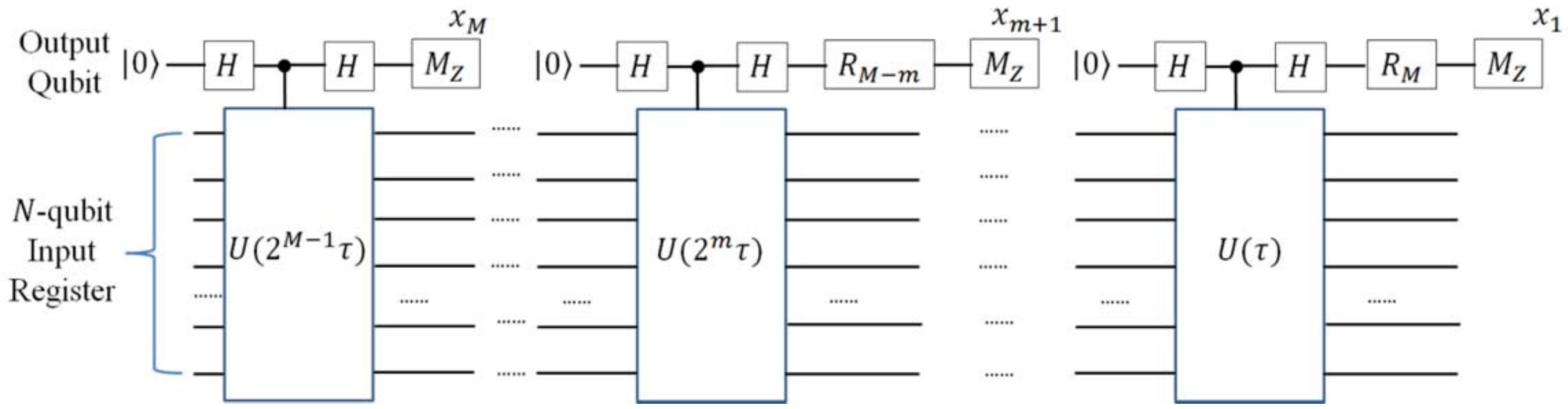

Fig. 6. Circuit for implementing the iterative phase estimation algorithm. Each qubit in the $N$-qubit input register corresponds to one of the $N$ spin-1/2 particles in the TIM model. Initially the $N$-qubit input quantum register is prepared with an estimate of the ground state of $H_I$. The output quantum register consists of a single-qubit recycled $M$ times. In each iteration, the output register gives the value of one bit in the $M$-bit binary expression of $\phi = 0.x_1 \dots x_M$. The Hadamard gate, the controlled unitary gate $U(2^m\tau)$, and the single qubit rotation gate $R_m$ are required. Details of the implementation can be found in Ref. [12].

Secondly, each gate in the phase estimation circuit is decomposed into a set of universal gates that can be implemented fault tolerantly with the surface code. The controlled-$U(2^m\tau)$ can be decomposed using the second-order Trotter formula [12,60,61] as illustrated in Fig. 7:

$$U(2^m\tau) = [U_x(\theta)U_{zz}(2\theta)U_x(\theta)]^k + \varepsilon_T, \quad (7)$$

where

$$U_x(\theta) = \prod_{j=1}^{N} u_j = \prod_{j=1}^{N} \exp(i\theta X_j/2), \quad (8)$$

$$U_{zz}(\theta) = \prod_{j=1}^{N-1} u_{jj+1} = \prod_{j=1}^{N} \exp(i\theta Z_j Z_{j+1}/2), \quad (9)$$

with $\theta = 2^m\tau/k$ and $\varepsilon_T$ the Trotter approximation error which can be made small by increasing the Trotter step $k$. $k$ is increased until $\varepsilon_T$ is less than $1/2^{M-m}$, which is the precision requirement of the controlled-$U(2^m\tau)$ gate. For a given $M$, a numerical value for the Trotter parameter $k(m=0) = k_0$ is found with the constraint $\varepsilon_T(m=0) < 1/2^M$. For $m > 0$, $k = 2^m k_0$ is set, satisfying the scaling of the error bound with $k$. The controlled-$U_x(\theta)$ and controlled-$U_{zz}(\theta)$ gates can be further decomposed into single-qubit rotations about the z axis, $R_z$, and CNOTs, shown in Figs. 8 and 9, respectively. Unlike the concatenation approach, additional qubits are not required to prepare an $N$-qubit cat state in order to parallelize each of the $N$ CNOT gates, because the surface code allows a single-control-bit multiple-target CNOT gate in the same amount of time as a single CNOT. The $R_z$ gates can be approximated using the set of basic gates $(H, S, T)$ by the Solovay-Kitaev theorem [62,63] (see the Appendix) with precision $\varepsilon_{sk}$. The limit on the Solovay-Kitaev error $\varepsilon_{sk} < \varepsilon/k$ is required in order that the total error in the approximation of $U(2^m\tau)$ is less than that required for the desired precision [12].

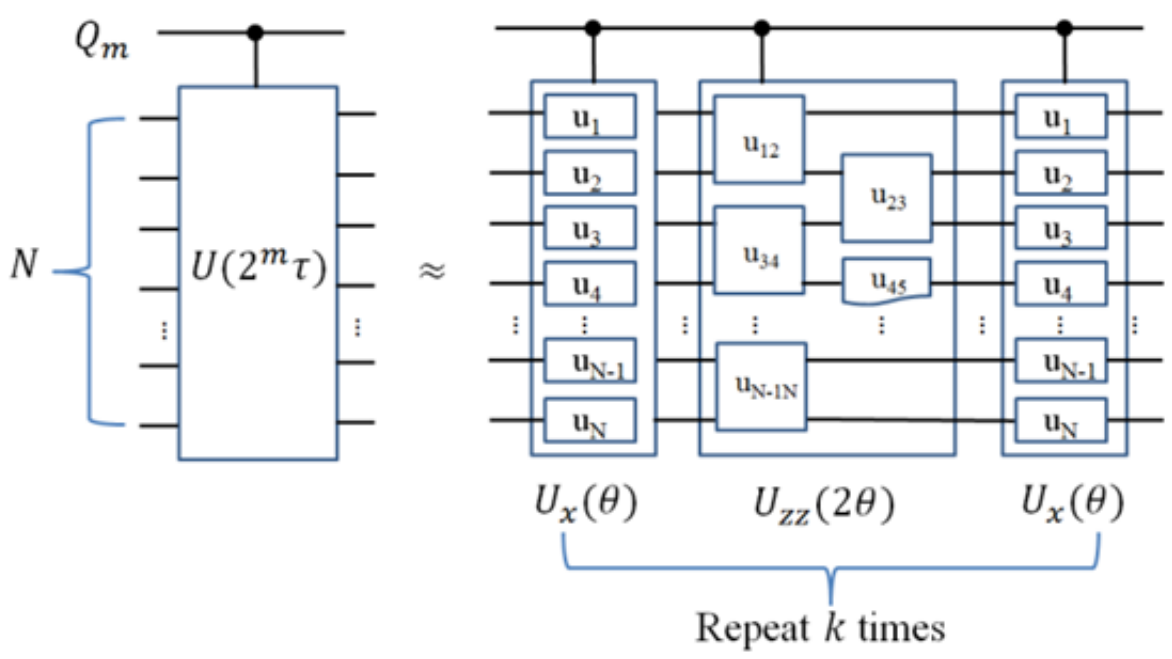

Fig. 7. Circuit for the controlled-$U(2^m\tau)$ gate approximated using the second-order Trotter formula. The resulting decomposed gates are the controlled- $U_x(\theta)$ and controlled- $U_{zz}(\theta)$ gates. The controlled-$U(2^m\tau)$ gate corresponds to the $(M-m)$th measured bit in the binary fraction for the phase $\phi$ and thus requires that the Trotter error must be less than the bit precision. Details of the implementation can be found in Ref. [12].

A complete decomposition of the controlled-$U(2^m\tau)$ into the set of basic gates is implemented on the surface code. From parameters in Table 1, the number of time steps $S_R$ (in units of surface code cycles) required to implement an $R_z$ gate is given by

$$S_R = 11.25dN_T + 10dN_S + 2.5dN_H, \quad (10)$$

where $N_T$, $N_S$, and $N_H$ are the numbers of T, S, and H gates respectively in the Solovay-Kitaev sequence approximating the $R_z$ gate. The number of time steps to implement the controlled-$U_x(\theta)$ and controlled-$U_{zz}(\theta)$ gates are $3S_R + 10d$ and $6S_R + 20d$, respectively. Each $R_j$ gate in Fig. 6 is equivalent to a rotation gate $R_z$ and requires $S_R$ basic gates. Thus the total number of time steps $K$ in surface code cycles required to implement the TIM circuit is given by

$$K = \sum_{m=0}^{M-1} 2^m k_0(9S_R + 30d) + 3S_R + 10d + S_R$$
$$= 2^{M-1}k_0(9S_R + 30d) + 4M(S_R + 2.5d). \quad (11)$$

$K$ therefore depends exponentially on the precision $M$. This is due to the fact that the Trotter parameter scales exponentially with $M$. A major difference with the concatenation code is that simulation time now scales linearly with code distance $d$ (compare with Eq. 9 of Ref. [12]).

As can be seen in Fig. 6, $N$ logical qubits are needed for the input register and one logical qubit is needed for the output register. We assumed that they are all smooth qubits in the surface code. The surface code implementation of a CNOT gate between two smooth qubits minimally requires a patch with 3 smooth logical qubits (the corresponding volume per unit time is 6) [38]. In addition, the surface code implementation of each T gate at most requires two auxiliary logical qubits: one qubit for preparation of the $|A_L\rangle$ state, another qubit for preparation of the $|Y_L\rangle$ state in case that $SX$ operation is required. Thus, in every surface code cycle, the number of logical qubits $Q$ required to implement the TIM simulation is $3(N+2)$.

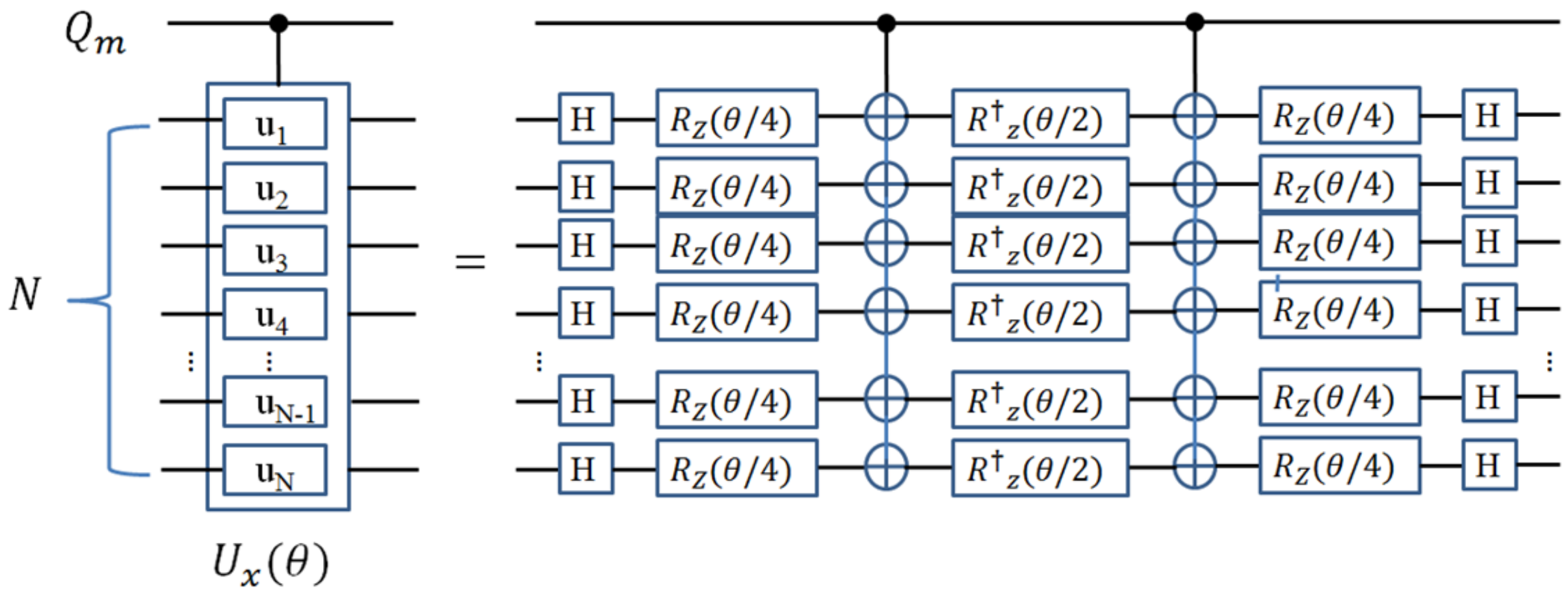

Fig. 8. The decomposition of the controlled-$U_x(\theta)$ into single-qubit $R_z$ gates and CNOT gates. There is a single-control-bit multiple-target CNOT gate in the circuit, which is different from the implementation in Ref. [12].

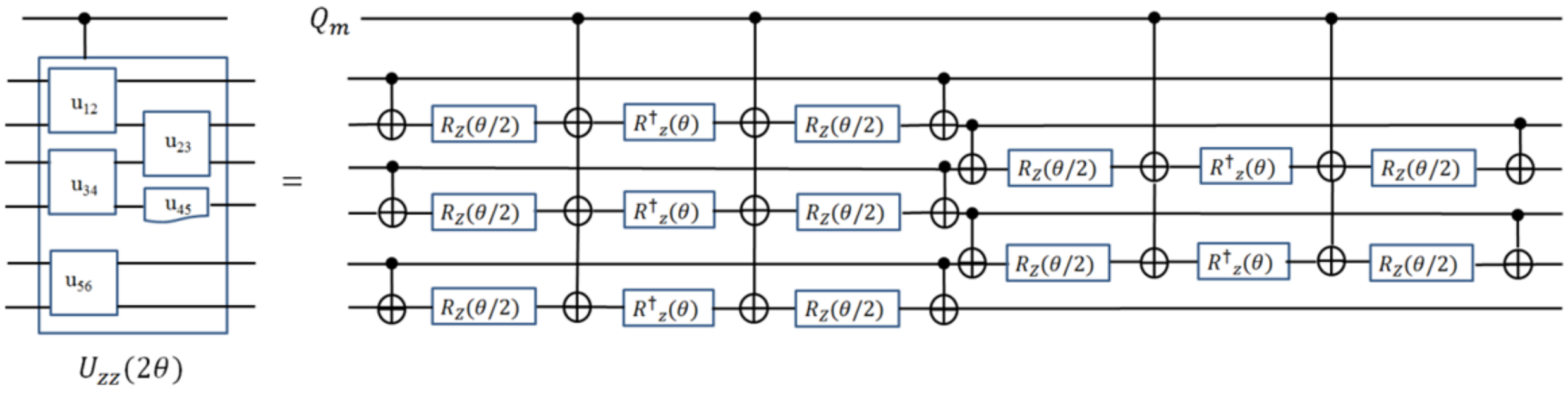

Fig. 9. The decomposition of the controlled-$U_{zz}(\theta)$ into single-qubit $R_z$ gates and CNOT gates. There is a single-control-bit multiple-target CNOT gate in the circuit, which is different from the implementation in Ref. [12].

Thirdly, we can estimate the strength of the error correction by the precision requirement of the simulation algorithm from the above results. The product $KQ$ is an upper bound on the total number of logical gates executed during the computation. The maximum failure probability in the entire quantum algorithm $p_t$ is given by

$$p_t = (KQ)p_L. \quad (12)$$

So, the logical error rate per surface code cycle $p_L$ must satisfy

$$p_L \ll 1/(KQ). \quad (13)$$

Given the ratio $p/p_{th}$, the values of $K$ and $Q$, one can deduce the code distance $d$ from Eqs. (1), (11) and (13) such that the failure probability of the entire quantum algorithm is sufficiently small. Since $d$ determines the spatial separations of the logical qubits and the temporal separations of the logical operations in the surface code, the array size of the surface code quantum computer and the implementation time are deduced from $d$.

To illustrate the resource requirements for the surface code, we adopt typical numerical values for the TIM problem. Let the number of spins $N = 100$ and $\frac{p}{p_{th}} = 0.1$ with the constraint that the maximum logical error rate per surface code cycle $p_L = r/(KQ)$. $r$, which is constrained to $0 < r \leq 1$, is a parameter related to the failure probability for the entire quantum algorithm. $r = 1$ corresponds to the worst case where the algorithm completes execution only 36% of the time [12]. Fig. 10 shows a plot of $d$, assuming $N = 100$, as a function of the desired maximum precision $M$. For a given value of $r$, $d$ increases as the precision increases, indicating that higher precision requires an increase in the size of the surface code quantum computer and longer computational times. Similarly, for a given $M$, $d$ increases as $r$ decreases. This is due to the fact that smaller $r$ increases the requirement for successful simulation runs in the algorithm. Such rigorous requirements result in a larger surface code quantum computer and increase the computational time.

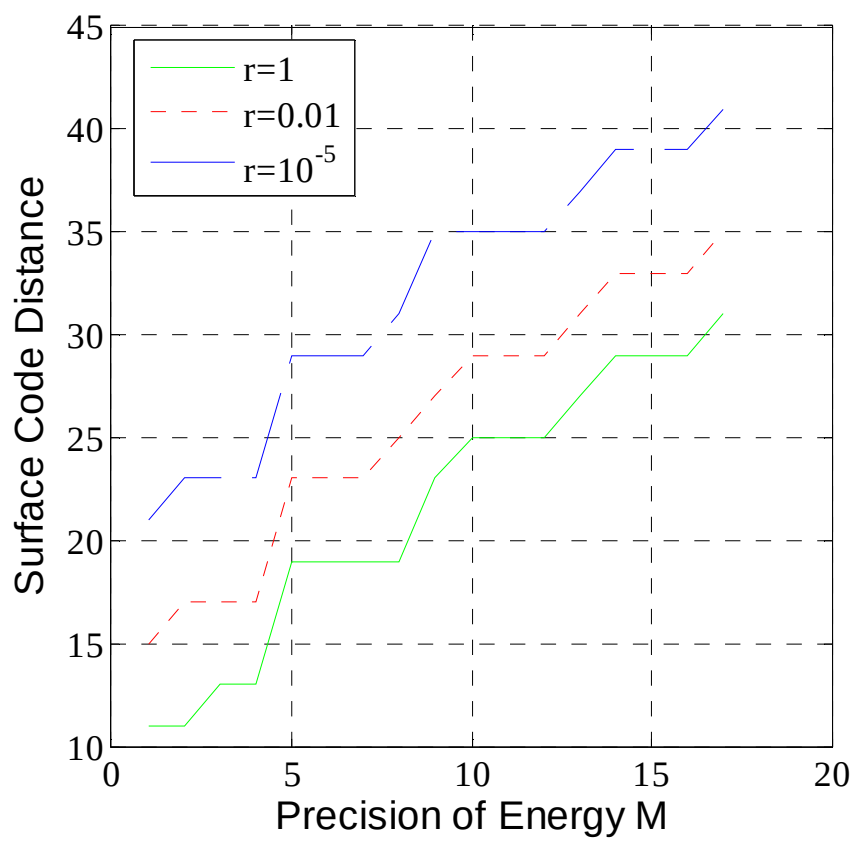

Fig. 10. Numerical simulation of the surface code distance $d$ assuming a $N = 100$ spin TIM problem, as a function of the desired maximum precision $M$.

The overhead of the surface code is mostly due to the ancilla requirement for the production of the T gates. As discussed above, the number of T gates in the algorithm is large and each T gate requires a high fidelity $|A_L\rangle$ state which can't be reused for another T gate. This $|A_L\rangle$ state is prepared by state distillation. Once the ancilla fidelity is higher than the necessary logical fidelity determined by Eq. (13), we may construct arbitrary fault-tolerant logical gates. We examine here the resource costs for this process. Numerical results show that level-2 distillation is sufficient for the parameters chosen for the TIM simulation. One might think that 225 logical qubits are required for the level-2 distillation process, because 225 approximate input states $\rho_A$ are prepared with the outputs of each distillation circuit providing the fifteen input states for the second distillation circuit. However, 225 logical qubits are actually excessive as the first level of the state distillation does not need to use the same amount of error correction as that of the second level [38]. It can easily use half the level 2 code distance and leads to a factor of 8 volume saving. As such, it would be better to study the resource cost of the distillation using the concept of volume [38].

Note that many state distillation procedures can occur in parallel enabling an arbitrary large number of $|A_L\rangle$ states to be generated in a given amount of time, depending on the desired qubit overhead.

As illustrated in Fig. 11, here we assume the most stringent situation for the implementation of $R_z$ gates using the Solovay-Kitaev sequence. The qubits representing $N$ spins in the TIM undergo T gates which are separated only by H gates. As the performance time of an H gate is shorter than T or S gates, this requires faster distillation speed of the $|A_L\rangle$ states than distillation speed requirements of other types of Solovay-Kitaev sequences. In this situation, we find that production of $N$ distilled $|A_L\rangle$ states in $13.75d$ surface code cycles (i.e., the duration time of a T gate and an H gate) is sufficient for the TIM simulation. Thus the required number of $|A_L\rangle$ states per unit time is $N\left(\frac{13.75d}{1.25d}\right)^{-1} = N/11$ (where the unit time is assumed to be $1.25d$surface code cycles). Considering that level-2 distillation is required and that the duration time for level-2 distillation is $6 + \frac{6}{2} = 9$ unit time, the required $|A_L\rangle$ state distillation factory volume per unit time should be $\frac{N}{11} \times \left(\frac{192}{6} + \frac{192\times15}{8\times3}\right) = 13.82N$, corresponding to a patch with $6.91N$ double defect smooth logical qubit overhead. Thus we would argue that this overhead is already a very acceptable overhead [38].

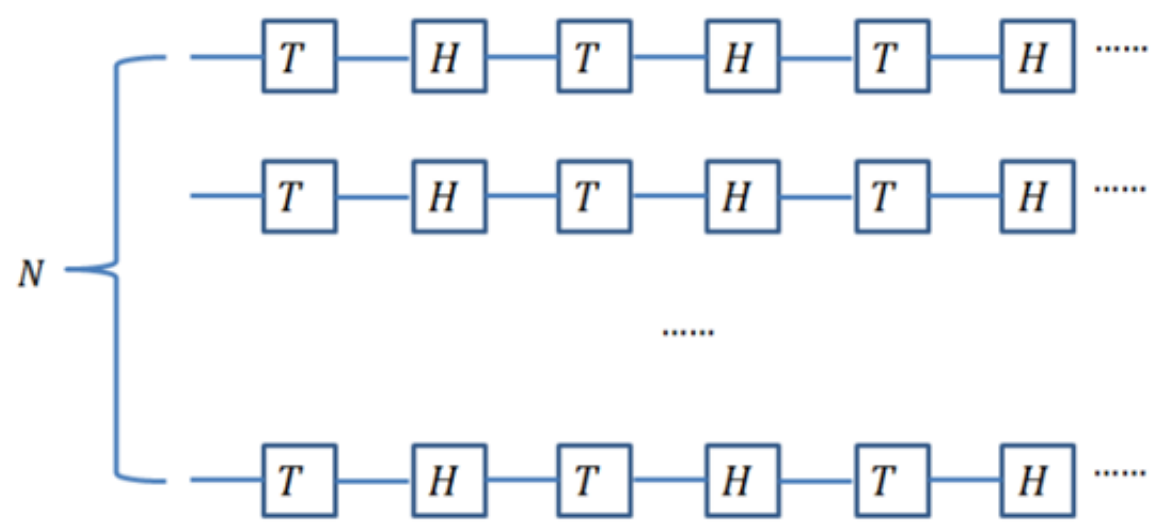

Fig. 11. The limiting factor on the distillation speed of the $|A_L\rangle$ states. As the quantum simulation is proceeding, the distillation of $|A_L\rangle$ occurs simultaneously. Thus the speed of the distillation must be sufficient for the consumption of $|A_L\rangle$ states in the simulation. That is, during the time period of two-level distillation, the number of the distilled $|A_L\rangle$ states consumed in the simulation must be equal to that of the $|A_L\rangle$ states produced in the distillation. Here we consider the situation where T gates are separated by only one H gate and thus require the fastest distillation rate.

After considering the overhead of the surface code, resource estimation can be made. Fig. 12 displays the number of physical qubits, assuming the $N = 100$ spin TIM problem ($r = 1$), as a function of the desired maximum precision $M$. A plot of the physical computation time as a function of the desired maximum precision $M$, assuming an average of 20 ns for each of the physical gates [36], is given in Fig. 13. As can be seen in the figure, the resources are large: for $m = 10$, $\sim 10^7$ physical qubits are required with a running time of $\sim 5$ hours.

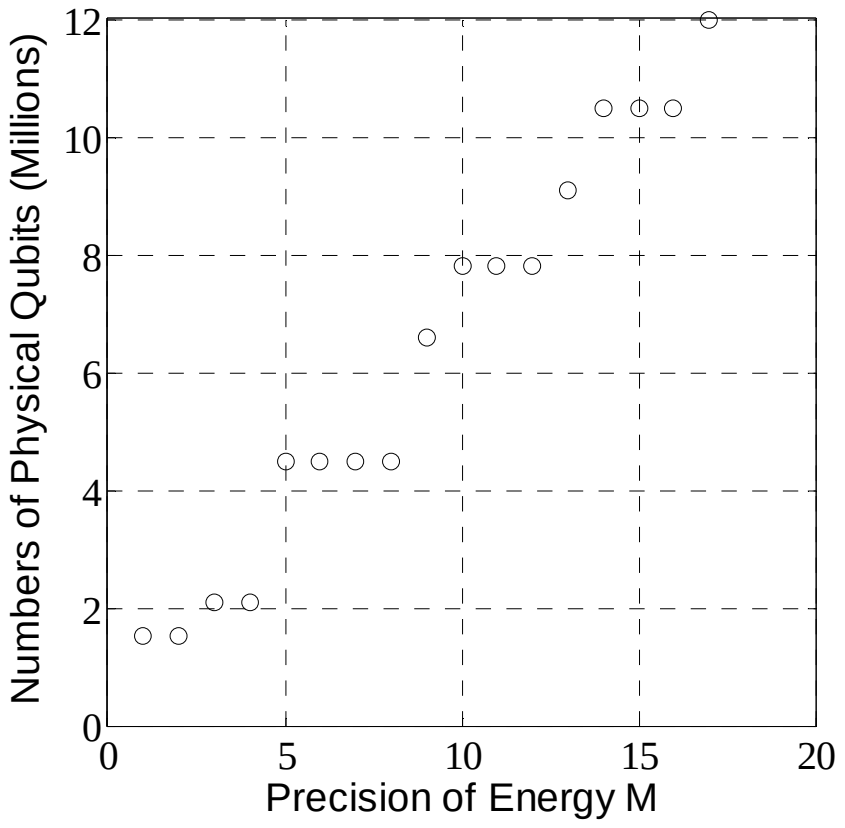

Fig. 12. Numerical simulations for the number of physical qubits, assuming a $N = 100$ spin TIM problem, as a function of the desired maximum precision $M$.

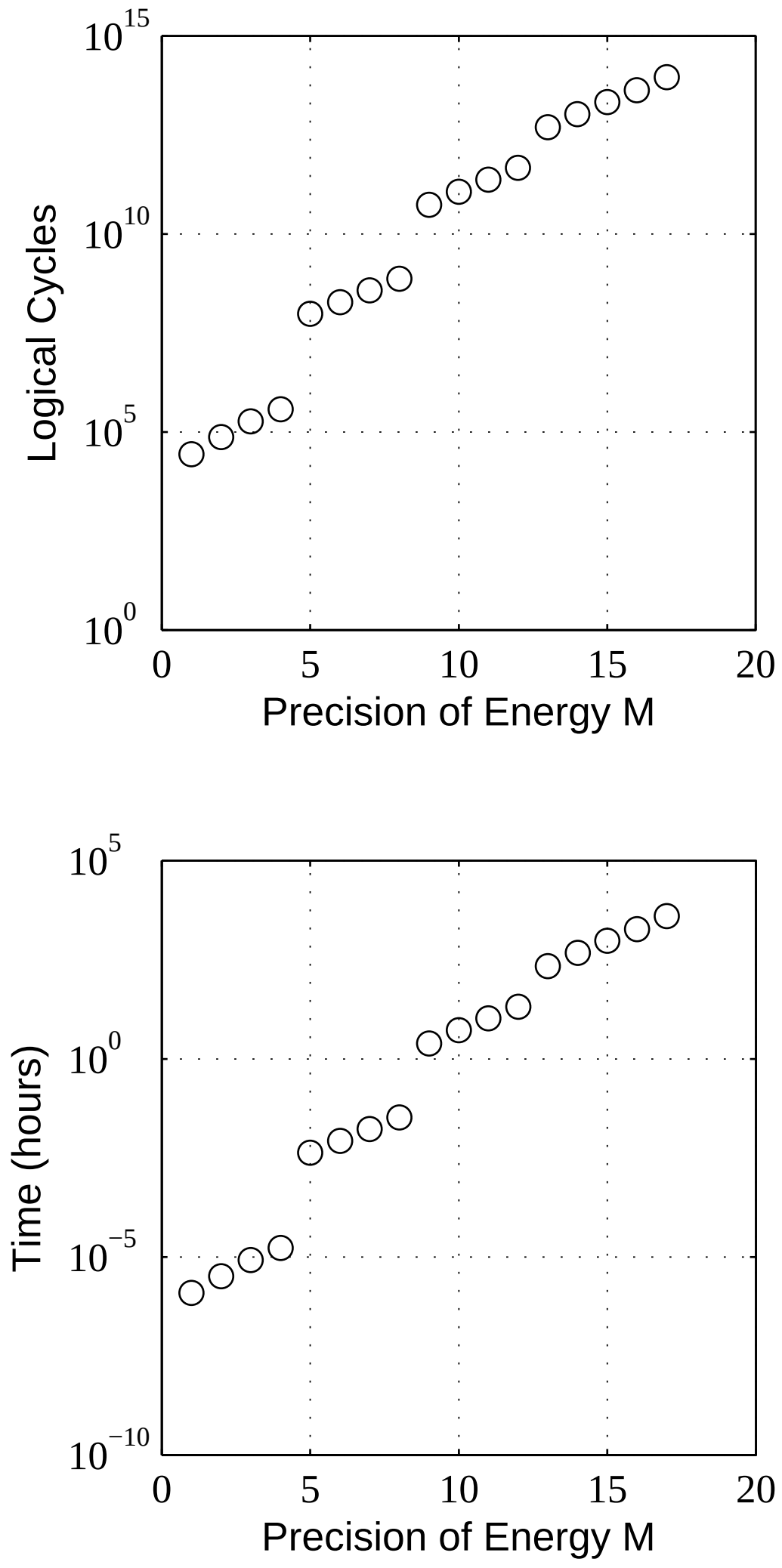

Fig. 13. Numerical simulations of the number of logical cycles (upper subfigure) and the number of physical time (lower subfigure), assuming the $N = 100$ spin TIM problem, as a function of the desired maximum precision $M$.

In order to compare with the results of the concatenation code implementation for error correction, an additional set of parameters are chosen as given in Ref. [12]: $\frac{p}{p_{th}} = 3\%$, physical gate time $\sim 10^{-5}$s for ion-trap devices, also with $N = 100$. One should notice that these parameters assumed very high fidelity gate performance $(1 \times 10^{-7})$ and a very high fault-tolerant threshold for the Steane code $(3.1 \times 10^{-6})$ which are likely overly optimistic for any currently available quantum computing architecture. Table 2 lists the resulting resource requirements for the surface code and the concatenation code results from Ref. [12]. The simulation time is comparable for $M \leq 18$. However, a higher precision requirement leads to higher levels of concatenation, which results in polynomially longer time for implementation, while the surface code implementation time scales as $d$ which is a logarithm dependence on the precision. For the physical qubits, the surface code requires 1-2 orders of magnitude more qubits than that of the concatenation code. However, this is a consequence of allowing for a significantly higher error threshold and is therefore more realistic in terms of current and near-term physical qubit performance.

| | Threshold | Time (days) | | Physical Qubits |
|---|---|---|---|---|
| Concatenation Code | $3.1 \times 10^{-6}$ | $M = 10$ | $\sim 10^2$ | $21^2 \times 4N \sim 10^5$ |
| | | $M = 16$ | $\sim 10^4$ | |
| Surface Code | 1% | $M = 10$ | $\sim 10^2$ | $9.91N \times 12.5d^2 \sim 10^6$ |
| | | $M = 16$ | $\sim 10^4$ | |

Table 2. Comparison of the performance of the surface code with that of the concatenation code [12].

## IV. SUMMARY AND CONCLUSIONS

We estimate the resource requirements for a quantum simulation of the ground state energy for the one-dimensional quantum transverse Ising model (TIM), based on the surface code implementation of error correction for a quantum computer. Comparing with previous results for the same model using the concatenation code error correction scheme, the current findings indicate that the simulation time is comparable, but that the number of physical qubits for the surface code are 1-2 orders of magnitude larger. Considering that the error threshold for the surface code is four orders of magnitude less restrictive than required in the concatenation code analysis, building a surface code quantum computer may be more feasible given current physical qubit performance parameters. Further, as a major contributor to the large resource requirements estimated here stem from the adoption of the Solovay-Kitaev theorem, utilization of more efficient decomposition schemes [64] may significantly reduce the resource cost. These improved decomposition schemes as well as high-order Trotterization will be explored in future work.

We would like to acknowledge valuable discussions with Joydip Ghosh, Matteo Mariantoni, Andrew Sornborger, James Whitfield and Zhongyuan Zhou. This work was supported by the National Science Foundation through grant CDI 1029764.

## APPENDIX. THE SOLOVAY-KITAEV ALGORITHM AND THE CONCATENATION CODE

The Solovay-Kitaev code utilized in Ref. [12] is undocumented while in this paper, the quantum compiler using the Solovay-Kitaev algorithm is adapted from an open source package [65]. The main goal of this paper is to illustrate the fundamental impact of the surface code on the quantum simulation of the TIM problem. The details of the Solovay-Kitaev code is not our primary focus. Nevertheless, for comparison purposes, we reproduce in this Appendix the numerical calculations for the number of logical cycles as a function of the desired precision in the simulation of the $N = 100$ spin TIM problem with the concatenation code, which was performed in Ref. [12]. (see Fig. 6 of Ref. [12]).

The total number of time steps $K$ required to implement the TIM quantum circuit using the concatenation code is given by (see Eq. 9 of Ref. [12]):

$$K = \sum_{m=0}^{M-1}[2^m k_0(9S_R + 11) + 4S_R + 4], \qquad \text{(A1)}$$

where $S_R$ is the length of the longest sequences of H, T, and S gates required to approximate the single unitary rotation gates using the Solovay-Kitaev algorithm. $M$ is the precision of the ground state energy. As can be seen in Fig. 14, no error correction is required for $M \leq 4$, since the maximum failure probability per gate $1/(KQ)$ is still below the physical ion-trap gate reliability $1 \times 10^7$. For $M \geq 5$, error correction is required which can explain a sudden jump in the number of time steps at $M = 5$. There is another jump in the number of time steps at $M = 9$, which is due to the increase of the Solovay-Kitaev order. Differences with the results reported in Ref. [12] are likely due to the adopted Solovay-Kitaev algorithm.

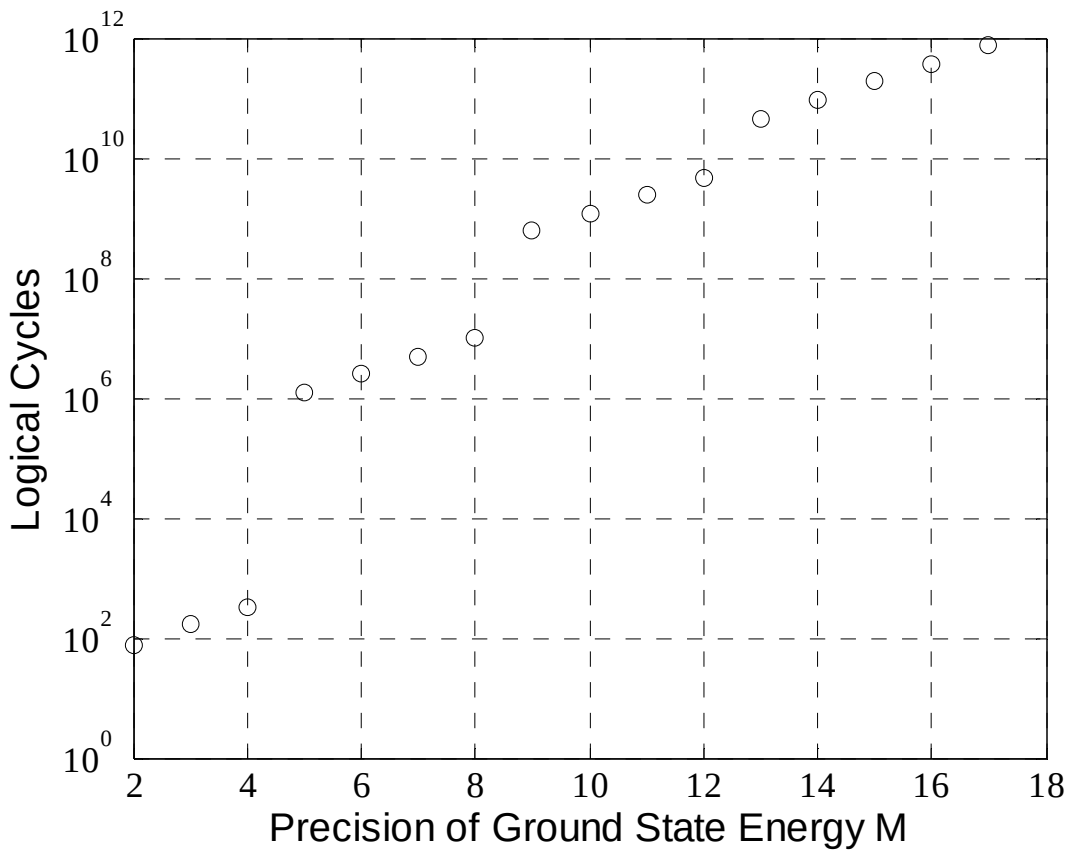

Fig. 14. The numerical calculations for the number of logical cycles $K$ required in the spin TIM simulation with the concatenation code as a function of the desired maximum precision $M$.